\documentclass{article}
\usepackage{authblk}
\usepackage[margin=.73in]{geometry}
\usepackage{amsmath}
\usepackage{graphicx}
\usepackage{caption}
\usepackage{subcaption}
\begin{document}

\title{Singularities in the acoustic Casimir pressure as a function of reflectivities}
\author{Thomas J. Liebau \\ (tliebau@vt.edu) }
\affil{\textit{Department of Biomedical Engineering and Mechanics, Virginia Tech, Blacksburg, VA 24060, USA.} \\ \textit{Ronin Institute, 127 Haddon Place, Montclair, NJ 07043, USA.}}

\maketitle

\noindent \textbf{Abstract}.  We study the acoustic Casimir pressure between imperfectly reflecting plates immersed in various isotropic noise backgrounds.  Unlike the case of perfect reflectors, the force tends to strong repulsion at small plate separations due to the existence of modes between the plates.  Additionally, in a noise field of constant spectral intensity, the large-bandwidth limit of the pressure has a reflectivity-dependent sensitivity to the infrared cutoff; we uncover the associated singular behavior as the product of the reflection coefficients approaches 1, which is intimately connected to a tendency of closely-spaced plates to repel if placed in a narrowly peaked spectral background.  

\section*{Introduction}

Two parallel plates in a random acoustic field will experience a force analogous to the quantum Casimir force [1].  Owing to the truncated bandwidth of the noise, this force oscillates between attraction and repulsion as a function of the plate separation $L$. Larraza et al. [2] proposed and experimentally demonstrated this acoustic Casimir effect, while also providing the theory for perfectly reflecting plates.  

If the plates are not perfectly reflective, the wavevector component perpendicular to the plates is no longer restricted to integer multiples of $\pi/L$, and the total density of modes must be calculated.  Esquivel-Sirvent et al. [3] derived an expression for the resulting pressure using a Green’s function approach.  In this work, we investigate subtle behavior of the acoustic Casimir force, at small plate separations or with large noise bandwidths, that does not occur for perfect reflectors.  

When perfectly reflecting plates are separated by less than the smallest half-wavelength contained in the background noise, the force is attractive and independent of the separation (no modes exist between the plates) [2].  This is no longer the case for imperfect reflectors, and the force is actually repulsive at sufficiently small plate separations.  Another contrast is that, in a background of constant spectral intensity $I_{\omega}$ over an infinite frequency range $(0,\infty)$, the force is technically zero; however, taking into account an infrared cutoff on the noise, we get results that tend smoothly to $-\pi I_{\omega}/4L$ as the product of the plate reflectivities, $\eta$, tends to 1.  Specifically, we derive an associated dimensionless cutoff, $\tilde{q}_{c} \sim (1-\eta)^{1/2}$, that quantifies the sensitivity of wideband limits to the length scale under consideration.

We then derive expressions for the pressure resulting from arbitrary power-law noise spectra.  These expressions are applied to the study of narrow unimodal spectra, which can model nonequilibrium fluid backgrounds [4].  The force between imperfect reflectors in a narrowly peaked noise background also tends rapidly to repulsion as the plate separation tends to zero; and in fact, the critical separation at which the sign of the force switches is closely related to the above dimensionless infrared cutoff pertaining to a flat spectral background. 

\section*{Flat spectrum}

\begin{figure}[h!]
  \centering
  \begin{subfigure}[b]{0.49\linewidth}
    \includegraphics[width=\linewidth]{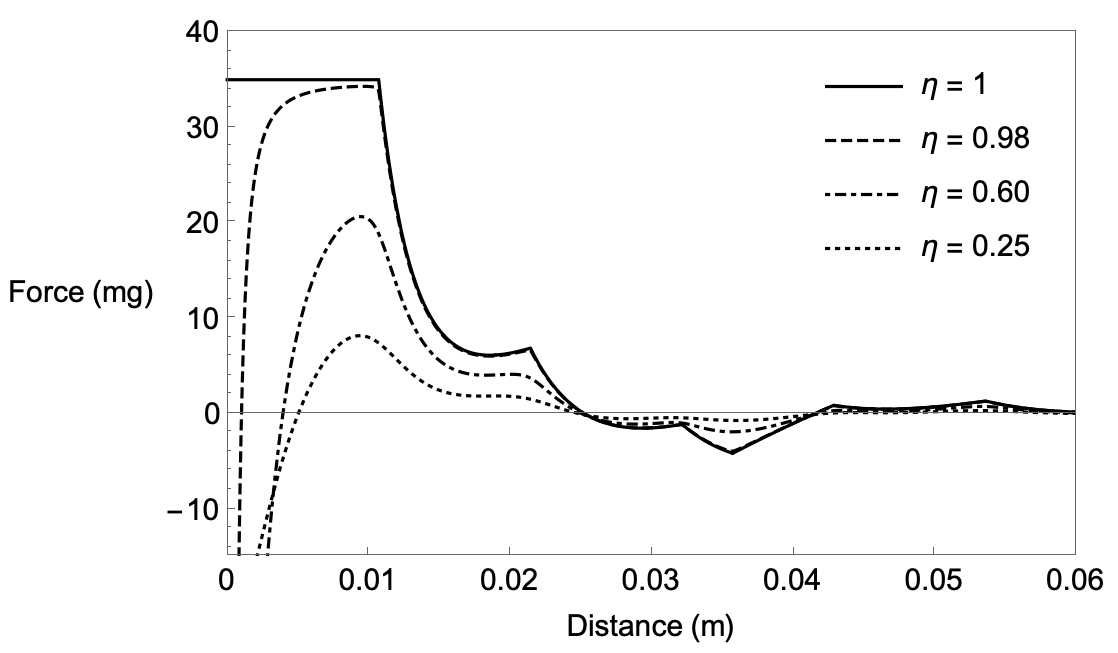}
    \caption{}
  \end{subfigure}
  \begin{subfigure}[b]{0.465\linewidth}
    \includegraphics[width=\linewidth]{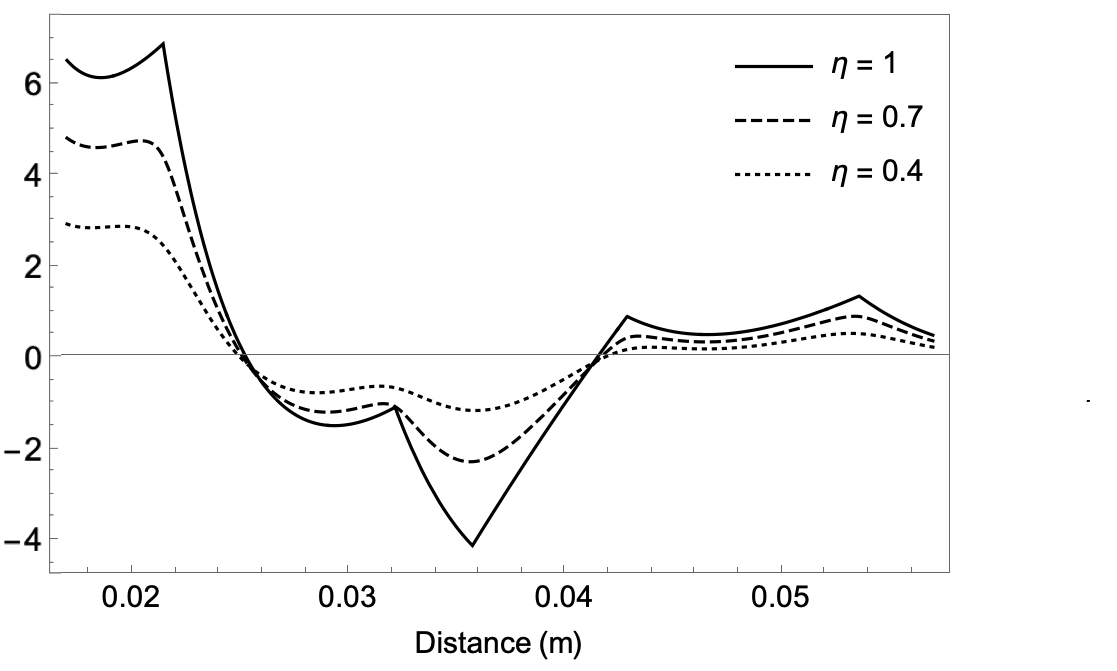}
    \caption{}
  \end{subfigure}
  \caption{Band-limited acoustic Casimir force between imperfect reflectors.  An attractive/repulsive force is given by a positive/negative value respectively.  Here $f_{1}=4.8\text{ kHz}$, $f_{2}=16\text{ kHz}$, $I_{\omega}=2.84\cdot10^{-4}\text{ J}/\text{m}^{2}$, $c=343\text{ m/s}$ and the area of each plate is $1.77\cdot10^{-2}\text{ m}^{2}$.}
  \label{fig1}
\end{figure}

Consider two parallel rigid plates with reflectivities $r_{A}, r_{B}$ separated by a distance $L$, and define $\eta \equiv r_{A}r_{B}$.  If the plates are immersed in an isotropic noise field of constant spectral intensity $I_{\omega}$ within the frequency range $[f_{1},f_{2}]$, the resulting net pressure is $\text{Re}[P]$, where [3]
\begin{equation}
P(\eta,L,q_{1},q_{2})=\frac{I_{\omega}}{2\pi}\iiint\limits_{q_{1}<|\textbf{k}|<q_{2}}dk_{x}dk_{y}dk_{z}\frac{k_{z}^{2}}{|\textbf{k}|^{4}}\frac{1}{\eta^{-1}e^{-2i(k_{z}+i0^{+})L}-1} 
\end{equation}
and $q_{1,2}=2\pi f_{1,2}/c$.  From now on, the positive imaginary infinitesimal is omitted unless needed.  If $\eta$ is constant over the considered bandwidth, this integral can be expressed in terms of polylogarithms [5]: for $|\eta|\leq1$, 
\begin{align}
P &= I_{\omega}\int_{0}^{\pi}d\phi\text{ }\sin{\phi}\int_{q_{1}}^{q_{2}}dq\frac{\cos^{2}\phi}{\eta^{-1}e^{-2iLq\cos{\phi}}-1} \nonumber \\ 
&=I_{\omega}\int_{-1}^{1}du\int_{q_{1}}^{q_{2}}dq\frac{u^{2}}{\eta^{-1}e^{-2iuLq}-1} \nonumber \\
&=I_{\omega}\int_{-1}^{1}du\text{ }\frac{iu}{2L}\left[\text{ln}(1-\eta e^{2iuLq_{2}})-\text{ln}(1-\eta e^{2iuLq_{1}})\right] \nonumber \\
&=\frac{I_{\omega}}{L}\left[F(Lq_{1},\eta)-F(Lq_{2},\eta)\right] 
\end{align}
where
\begin{equation}
F(\tilde{q},\eta) = \frac{\text{Li}_{2}(\eta e^{-2i\tilde{q}})}{4\tilde{q}}+\frac{\text{Li}_{2}(\eta e^{2i\tilde{q}})}{4\tilde{q}}+\frac{\text{Li}_{3}(\eta e^{-2i\tilde{q}})}{8i\tilde{q}^{2}}-\frac{\text{Li}_{3}(\eta e^{2i\tilde{q}})}{8i\tilde{q}^{2}},
\end{equation}
and
\begin{equation*}
\text{Li}_{m}(z)\equiv\sum_{\nu=1}^{\infty}\frac{z^{\nu}}{\nu^{m}}, \ \ \ \ \ z\leq1.
\end{equation*}

In Fig. 1, we plot $-P(L)$ with fixed noise parameters for different values of $\eta$.  As $\eta\to1$, the result converges to that of Larraza et al. [2].  For sufficiently large plate separation, the main action of reduced reflectivity, besides reducing the magnitude of the force, is the smoothing of the kinks at which $q_{1,2}L/\pi$ is an integer.  These kinks are reproduced when $\eta=1$, owing to the branch point of $\text{Li}_{m}(z)$ at $z=1$.  

In this example, the sign of the force (as a function of $L$) is barely affected by $\eta$ for $L>\pi/q_{2}$.  For $L<\pi/q_{2}$, however, the behavior differs remarkably from the $\eta=1$ case; the force is no longer independent of the separation, and actually tends to repulsion.  In the limit of no separtion between the plates,
\begin{equation*}
\lim_{L\to0}P(1,L) = I_{\omega}\cdot\frac{q_{1}-q_{2}}{3},
\end{equation*} 
but if $\eta\neq1$,
\begin{equation}
\lim_{L\to0}P(\eta,L) = I_{\omega}\cdot\frac{2\eta}{3}\left(\frac{q_{1}-q_{2}}{\eta-1}\right).
\end{equation} 

Furthermore, in the double-limit $q_{1}\to0$ and $q_{2}\to\infty$, we have $P\to-\pi I_{\omega}/4L$ when the plates are perfect reflectors (corresponding to an attractive force) [2], but $P$ actually goes to \textit{zero} for any $\eta\neq1$.  This is demonstrated by plotting $-F(\tilde{q})$ with various $\eta$ in Fig. 2; when $\eta$ is slightly less than 1, $F$ has a minimum at some small $\tilde{q}_{c}$, with $F(\tilde{q}_{c})$ slightly greater than $-\pi/4$.  As $\tilde{q}$ is varied from $\tilde{q}_{c}$ to 0, $F$ rapidly rises to zero.

\begin{figure}[h!]
  \centering
  \begin{subfigure}[b]{0.49\linewidth}
    \includegraphics[width=\linewidth]{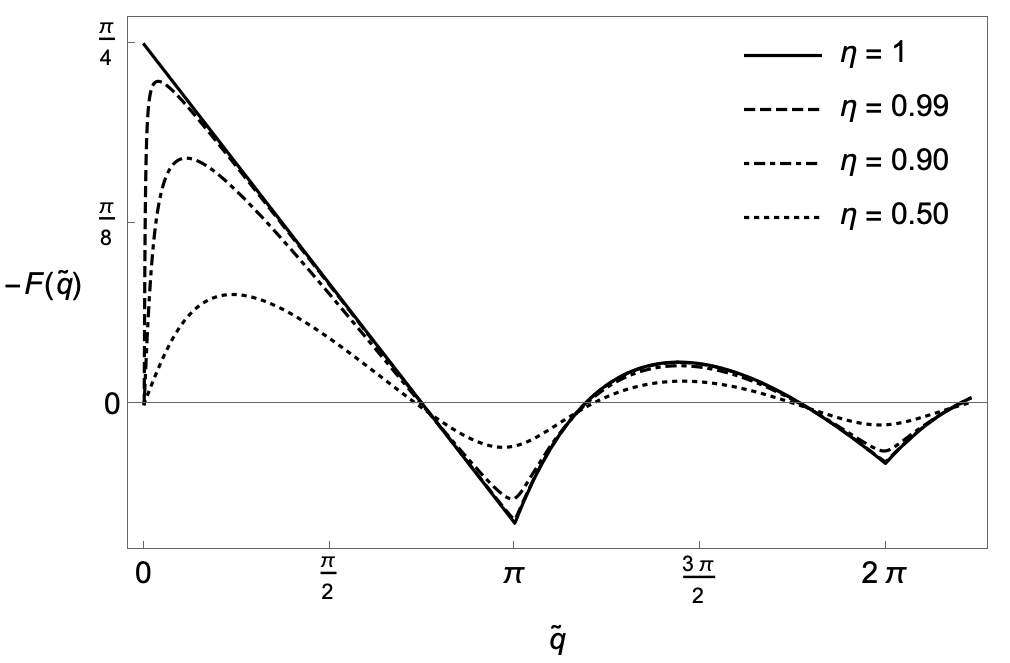}
    \caption{}
  \end{subfigure}
  \begin{subfigure}[b]{0.50\linewidth}
    \includegraphics[width=\linewidth]{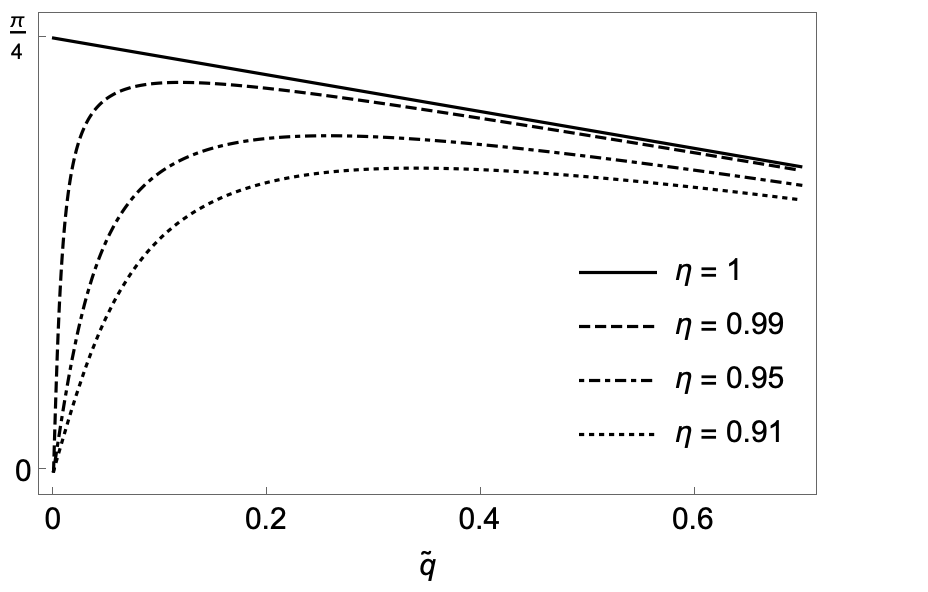}
    \caption{}
  \end{subfigure}
  \caption{The singular bahaviour of $F(\tilde{q},\eta)$ as $\eta$ approaches $1$.}
  \label{fig2}
\end{figure}

To resolve the issue of infinite-bandwidth limits, we consider $\tilde{q}_{c}(\eta)$ as a nondimensional infrared cutoff, and study its behavior as $\eta\to1$.  We define 
\begin{equation*}
\psi(\tau,\tilde{q}) \equiv  \left.\frac{\partial F}{\partial\tilde{q}}\right|_{\eta\to 1-\tau} \ \ \ \ \
\psi_{0}(\tilde{q}) \equiv  \psi(0,\tilde{q}) \ \ \ \ \ \ \
\psi_{1}(\tilde{q}) \equiv  -\left.\frac{\partial\mathcal{\psi}}{\partial\tau}\right|_{(0,\tilde{q})}
\end{equation*}
where $\tau\equiv 1-\eta>0$ plays a role analogous to a reduced temperature, and now 
\begin{equation*}
\tilde{q}_{c} = \min\{\tilde{q}\text{ }|\text{ }\psi(\tau,\tilde{q})=0\text{ \& }\tilde{q}>0\}.
\end{equation*}
As $\tau\to 0$, we have to lowest order in $\tilde{q}_{c}$,
\begin{equation}
\psi(\tau,\tilde{q}_{c}) = \psi_{0}(\tilde{q_{c}})-\tau\psi_{1}(\tilde{q_{c}}) = 0 \implies \tau = \frac{\psi_{0}(\tilde{q_{c}})}{\psi_{1}(\tilde{q_{c}})} = \frac{2\tilde{q_{c}}^{2}}{3} \Leftrightarrow \boxed{\tilde{q_{c}}(\eta) = \sqrt{\tfrac{3}{2}}(1-\eta)^{1/2}}
\end{equation}
and
\begin{equation}
F\left(\text{ }\tilde{q}_{c}\text{ },\text{ }1-\tfrac{2}{3}\tilde{q}_{c}^{2}\text{ }\right)=\frac{2\tilde{q}_{c}}{3}-\frac{\pi}{4} \Leftrightarrow \boxed{F(\tilde{q}_{c}(\eta),\eta) = \sqrt{\tfrac{2}{3}}(1-\eta)^{1/2}-\frac{\pi}{4}}
\end{equation}
This analysis is confirmed by numerically solving for the first zero of $\psi(\tau,\tilde{q})$ without taking any expansion.  We find that over the range $0.9<\eta<1$, the relations
\begin{equation}
\tau = \frac{2\tilde{q}_{c}^{2}}{3}+0.349\tilde{q}_{c}^{3}
\end{equation}
and
\begin{equation}
F\left(\text{ }\tilde{q}_{c}\text{ },\text{ }1-\tfrac{2}{3}\tilde{q}_{c}^{2}-0.349\tilde{q}_{c}^{3}\text{ }\right)=\frac{2\tilde{q}_{c}}{3}+0.0872\tilde{q}_{c}^{2}-\frac{\pi}{4} 
\end{equation}
are accurate to less than $1\%$ relative error. 

\begin{figure}[h!]
  \centering
  \begin{subfigure}[b]{0.49\linewidth}
    \includegraphics[width=\linewidth]{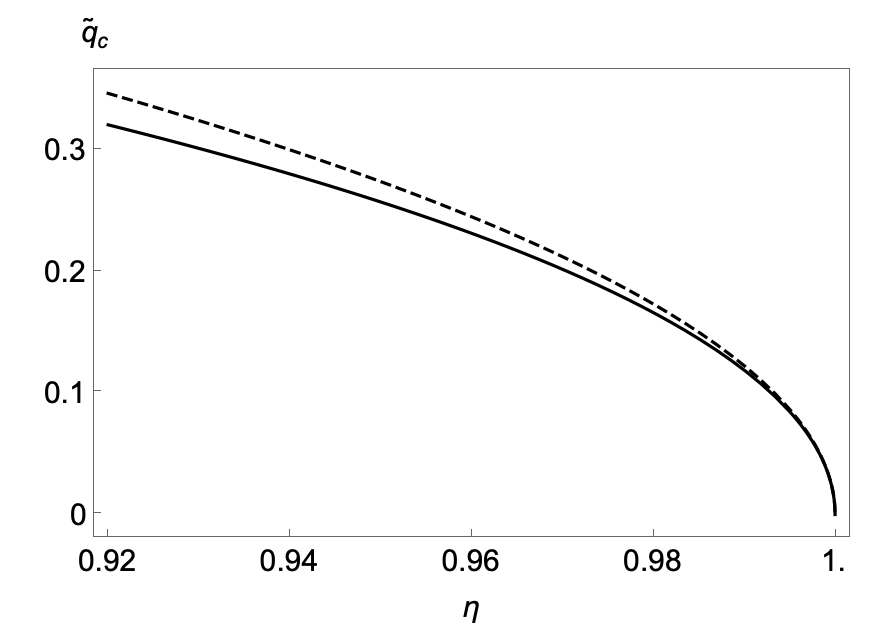}
    \caption{}
  \end{subfigure}
  \begin{subfigure}[b]{0.50\linewidth}
    \includegraphics[width=\linewidth]{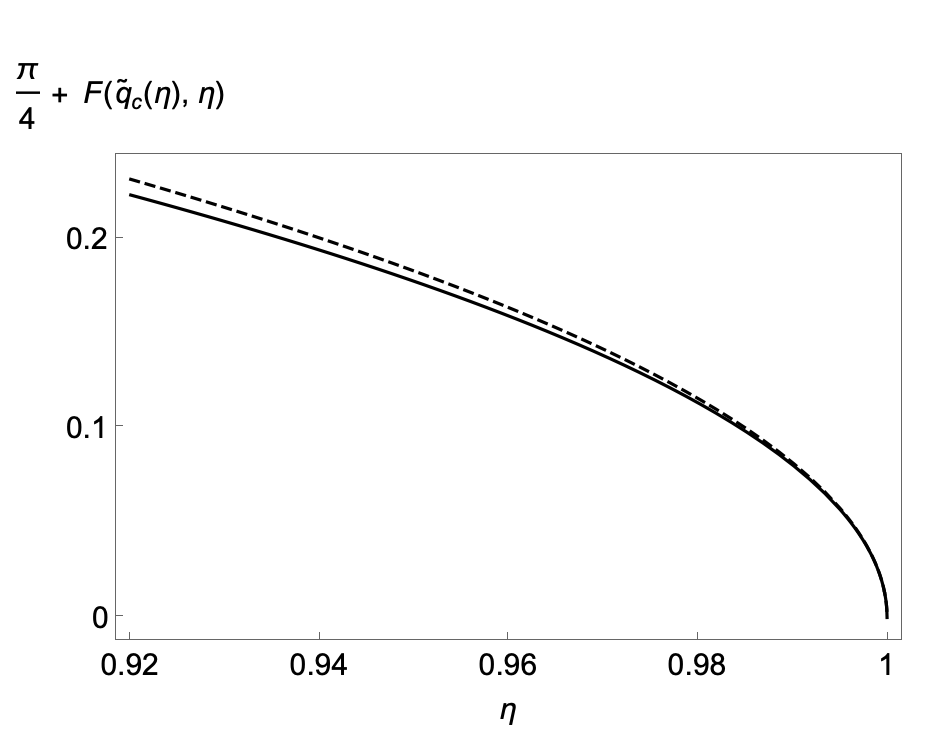}
    \caption{}
  \end{subfigure}
  \caption{(a) The cutoff, and (b) corresponding difference between $F$ and $-\pi/4$ as $\eta\to1$.  The solid lines are given by Eqs. (7) and (8) and are visually indistinguishable from numerical solutions.  The dashed lines are given by Eqs. (5) and (6). }
  \label{fig3}
\end{figure} 

Physically, a flat spectrum must have an infrared cutoff, which we can write as $q_{c} = \tilde{q}_{c}(\eta)/L_{0}$ for some length $L_{0}$.  The corresponding pressure is then
\begin{equation*}
\lim_{q_{2}\to\infty}P\left(\eta,L,\tfrac{\tilde{q}_{c}(\eta)}{L_{0}},q_{2}\right) \equiv P^{(q_{c},\infty)}(\eta,L) =\frac{I_{\omega}}{L}F\left(\tilde{q}_{c}(\eta)\tfrac{L}{L_{0}},\text{ }\eta\right), 
\end{equation*}
with $F$ given by Eq. (3).  If $\eta$ is near 1, then $P^{(q_{c},\infty)}(\eta,L_{0})$ is slightly greater than $-\pi I_{\omega}/4L_{0}$, cf. Fig. 3b.  Furthermore, since $\tilde{q}_{c}$ is small, $P^{(q_{c},\infty)} \approx -\pi I_{\omega}/4L$ over a large range of $L/L_{0}>1$, up to the point where $F\left(\tilde{q}_{c}(\eta)\tfrac{L}{L_{0}}\right)$ becomes too small; perhaps around $\tilde{q}_{c}L/L_{0} = \pi/4$, cf. Fig 2a.  At around $\tilde{q}_{c}L/L_{0} = 3\pi/4$, the force becomes repulsive, and slowly oscillates between attractive and repulsive as $L/L_{0}$ increases further.  

The above analysis details the sensitivity of the acoustic Casimir force, in wideband limits, to the infrared cutoff, the reflectivities of the plates, and the considered length scales of their separation.  On the other hand, for a fixed wavenumber band with $q_{1},q_{2}$ of the same order, a similar analysis reveals that $L_{c}(\eta)$, the smallest plate separation at which the force is zero, is about $\sqrt{\frac{3(1-\eta)}{2q_{1}q_{2}}}$.  In the example of Fig. 1, this expression correctly predicts a critical separation of about $1\text{ mm}$ when the product of the reflectivities is 0.98.

\section*{Power-law and narrow unimodal spectra}

Next, we consider background spectra of the form $G(q) = G_{\alpha}q^{\alpha}$ over a wavenumber range $[q_{1},q_{2}]$, where $G_{\alpha}$ is a constant of dimension Energy/Length$^{2-\alpha}$.  The corresponding Casimir pressure is  $G_{\alpha}\text{Re}[P_{\alpha}]$, where
\begin{align}
P_{\alpha}(\eta,L,q_{1},q_{2})&=\int_{q_{1}}^{q_{2}}dq\text{ }q^{\alpha}\int_{-1}^{1}du\frac{u^{2}}{\eta^{-1}e^{-2iuLq}-1}  \\
&=\int_{q_{1}}^{q_{2}}dq\text{ }q^{\alpha}\left[f(Lq)+f(-Lq)\right], \nonumber
\end{align}
\begin{equation}
f(\tilde{q}) \equiv \frac{\text{ln}(1-\eta e^{-2i\tilde{q}})}{2i\tilde{q}}+\frac{\text{Li}_{2}(\eta e^{-2i\tilde{q}})}{2\tilde{q}^{2}}+\frac{\text{Li}_{3}(\eta e^{-2i\tilde{q}})}{4i\tilde{q}^{3}}.
\end{equation}
Since $|\eta|\leq1$, we can express $f$ in series form and carry out the $q$-integration term-by-term.  The result is
\begin{align}
P_{\alpha}(\eta,L,q_{1},q_{2})&= \sum_{\nu=1}^{\infty}\eta^{\nu}\int_{q_{1}}^{q_{2}}dq\left[e^{-2i\nu Lq}\left(\frac{q^{\alpha-3}}{4i\nu^{3}L^{3}} + \frac{q^{\alpha-2}}{2\nu ^{2}L^{2}} - \frac{q^{\alpha-1}}{2i\nu L}\right)+\text{c.c.}\right] \nonumber \\
&=\frac{1}{L^{1+\alpha}}\left[F_{\alpha}(Lq_{1},\eta)-F_{\alpha}(Lq_{2},\eta)\right],
\end{align}
where
\begin{equation}
F_{\alpha}(\tilde{q},\eta) = \sum_{\nu=1}^{\infty}\eta^{\nu}\left[\frac{\tilde{q}^{\alpha-2}}{4i\nu^{3}}E_{3-\alpha}(2i\nu\tilde{q})+\frac{\tilde{q}^{\alpha-1}}{2\nu^{2}}E_{2-\alpha}(2i\nu\tilde{q})-\frac{\tilde{q}^{\alpha}}{2i\nu}E_{1-\alpha}(2i\nu\tilde{q}) + \text{c.c.} \right]
\end{equation}
and
\begin{equation*}
E_{m}(z) \equiv \int_{1}^{\infty}dt\text{ }\frac{e^{-zt}}{t^{m}}.
\end{equation*}
This result, like that of the flat-spectrum case, separates the $L$ and $Lq_{1,2}$ dependencies of the pressure.  The large-$L$ asymptotics are given by
\begin{align*}
&\frac{F_{\alpha}(Lq,\eta)}{L^{1+\alpha}} \to \frac{q^{\alpha-1}}{L^{2}}\sum_{\nu=1}^{\infty}\frac{\eta^{\nu}}{\nu^{2}}\cos{(2\nu Lq)} = \frac{q^{\alpha-1}}{2L^{2}}\left[\text{Li}_{2}\left(\eta e^{2i\nu Lq}\right)+\text{Li}_{2}\left(\eta e^{-2i\nu Lq}\right)\right],
\end{align*}
\begin{figure}[h!]
  \centering
  \begin{subfigure}[b]{0.49\linewidth}
    \includegraphics[width=\linewidth]{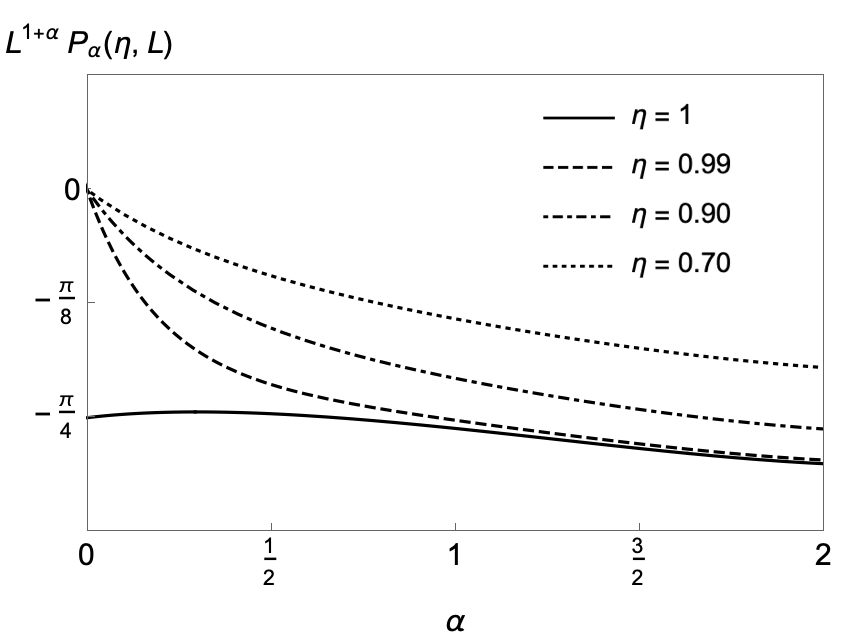}
    \caption{}
  \end{subfigure}
  \begin{subfigure}[b]{0.48\linewidth}
    \includegraphics[width=\linewidth]{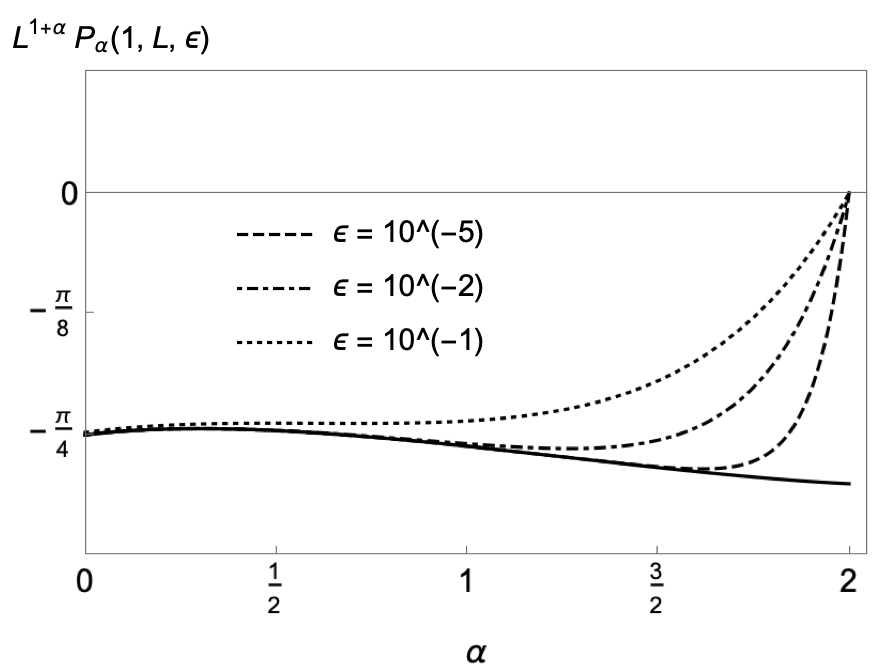}
    \caption{}
  \end{subfigure}
  \caption{Infinite-bandwidth limits for power-law backgrounds of exponent $\alpha$, (a) for various reflectivities with no angle cutoff, (b) for perfect reflectors with different oblique-angle cutoffs $\phi_{\varepsilon}=\arccos{\varepsilon}$.  The solid line in (b) gives the case $\varepsilon\to0$ and tends to $-\tfrac{\pi}{4}\zeta(3)$ as $\alpha\to2$, but for any nonzero $\varepsilon$ the fluctuation-induced force is zero when $\alpha=2$.}
  \label{fig4}
\end{figure}
revealing an oscillating $L^{-2}$ decay for an arbitrary band-limited spectrum.  Meanwhile, infinite-bandwidth limits are obtained for $0<\alpha<2$ as
\begin{align}
P_{\alpha}^{(0,\infty)}(\eta,L) &= \int_{-1}^{1}du\text{ }u^{2}\int_{0}^{\infty}dq\frac{q^{\alpha}}{\eta^{-1}e^{-2i(qu+i0^{+})L}-1}  \nonumber \\
&= \int_{0}^{1}du\text{ }u^{2}\int_{0}^{\infty}dq\left[\frac{q^{\alpha}}{\eta^{-1}e^{-2i(qu+i0^{+})L}-1} + \frac{q^{\alpha}}{\eta^{-1}e^{2i(qu-i0^{+})L}-1}\right] \nonumber \\
&= -\frac{2^{-\alpha}}{L^{1+\alpha}}\Gamma(1+\alpha)\sin{\left(\frac{\alpha\pi}{2}\right)}\text{Li}_{1+\alpha}(\eta) \int_{0}^{1}\frac{du}{u^{\alpha-1}} \\
&= -\frac{2^{-\alpha}}{L^{1+\alpha}}\frac{\Gamma(1+\alpha)}{2-\alpha}\sin{\left(\frac{\alpha\pi}{2}\right)}\text{Li}_{1+\alpha}(\eta),
\end{align}
where $\Gamma$ is the gamma function.  Eq. (13) is obtained by mapping the $q$-integration of the first (second) term in brackets to the positive (negative) imaginary axis.  For perfect reflectors, (14) reduces to
\begin{equation}
P_{\alpha}^{(0,\infty)}(1,L) = \left(\frac{\pi}{L}\right)^{1+\alpha}\frac{\zeta(-\alpha)}{2-\alpha},
\end{equation}
where $\zeta$ is the Riemann zeta function, recovering the result $-\pi/4L$ for a flat spectrum, $\alpha\to0$.  

The case $\alpha=2$ and $[q_{1},q_{2}]\to[0,\infty)$ corresponds to two plates immersed in a fluid at thermal equilibrium (neglecting an ultraviolet cutoff), $G(q)\propto q^{2}$, and we expect that the fluctuation-induced force would vanish [4].  However, (14) implies that as $\alpha\to 2^{-}$,
\begin{equation*}
P_{\alpha}^{(0,\infty)}(\eta,L) \to -\frac{\pi}{4L^{3}}\text{Li}_{3}(\eta).
\end{equation*}
To resolve this, we note that from (13),
\begin{align}
G_{\alpha}\text{Re}\left[P_{\alpha}^{(0,\infty)}\right] = -\int_{0}^{\pi/2}\frac{d\phi\text{ }\sin{\phi}}{(\cos{\phi})^{\alpha-1}}\frac{\Gamma(1+\alpha)}{(2L)^{1+\alpha}}\sin{\left(\frac{\alpha\pi}{2}\right)}\left[\text{Li}_{1+\alpha}^{}(\eta)+\text{Li}_{1+\alpha}^{}(\eta^{*})\right]G_{\alpha},
\end{align}
which gives the following physical picture:  The integrand of (16) is a ``pressure per radian,'' due to all waves that strike the plates between angles $(\phi,\text{ }\phi+d\phi)$ from an axis perpendicular to the plates.  This is true for \textit{any} $\alpha>0$; but, for a power-law spectrum of exponent near or greater than 2, the \textit{total} pressure is very sensitive to the waves coming in at oblique angles, and the integral over all angles is divergent for non-even $\alpha>2$.  Nonetheless, in an equilibrium fluid -- or in any background of positive even exponent -- the pressure that results from all the waves which strike the plate at a particular angle is zero.

\begin{figure}[h!]
  \centering
  \begin{subfigure}[b]{0.47\linewidth}
    \includegraphics[width=\linewidth]{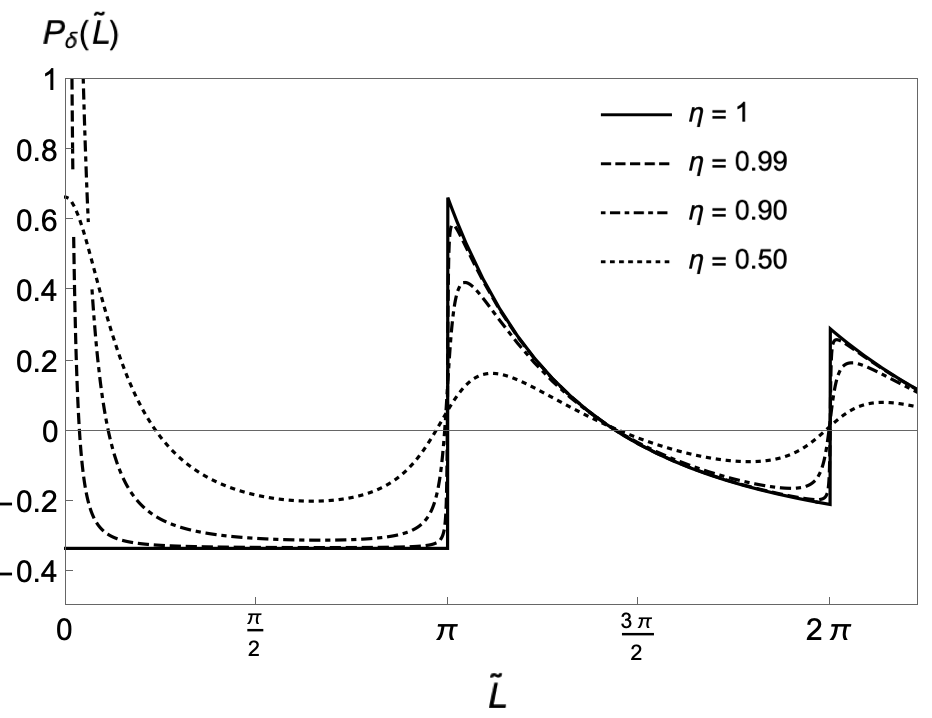}
    \caption{}
  \end{subfigure}
  \begin{subfigure}[b]{0.51\linewidth}
    \includegraphics[width=\linewidth]{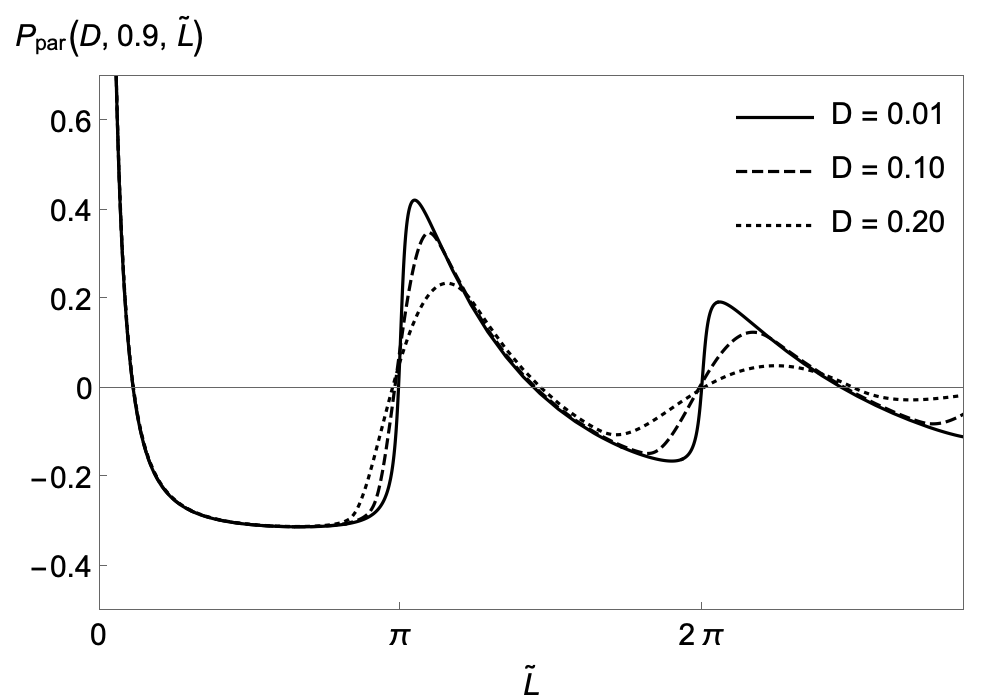}
    \caption{}
  \end{subfigure}
  \caption{Casimir pressure resulting from narrowly peaked spectra.  (a) Delta-function spectrum for various $\eta$.  (b) Parabolic spectrum for various peak widths with $\eta=0.9$.  The prediction of rapid variation near $\tilde{L}=\tilde{q}_{c}(\eta)$ is confirmed.  When $D=0.01$, the corresponding pressure is indistinguishable from $P_{\delta}(0.9,\tilde{L})$ over this range, but this would not be the case for larger $\tilde{L}$.}
  \label{fig5}
\end{figure}

This sensitivity should be investigated in cases of finite bandwidth.  If incident angles $\phi_{\varepsilon}<\phi<\tfrac{\pi}{2}$ are excluded, the total pressure is $G_{\alpha}$ times the real part of
\begin{align*}
&P_{\alpha}(\eta,L,q_{1},q_{2}) - \int_{q_{1}}^{q_{2}}dq\text{ }q^{\alpha}\int_{-\varepsilon}^{\varepsilon}du\frac{u^{2}}{\eta^{-1}e^{-2iuLq}-1} \nonumber \\ &= \frac{1}{L^{1+\alpha}}\left[F_{\alpha}^{}(Lq_{1},\eta)-F_{\alpha}^{}(Lq_{2},\eta)\right] -\frac{\varepsilon^{2-\alpha}}{L^{1+\alpha}}\left[ F_{\alpha}^{}(\varepsilon Lq_{1},\eta)-F_{\alpha}^{}(\varepsilon Lq_{2},\eta) \right],
\end{align*}
where $\varepsilon\equiv\cos{\phi_{\varepsilon}}\ll1$.  This would seem like a drastic deviation from the result (11)-(12) when $\alpha$ is near or greater than 2; but, noting that the oscillating function $F_{\alpha}(x)$ grows in magnitude like $x^{\alpha-1}$, the quantity $\varepsilon^{2-\alpha}\left[F_{\alpha}^{}(\varepsilon\tilde{q}_{1})-F_{\alpha}^{}(\varepsilon\tilde{q}_{2})\right]$ is roughly $\mathcal{O}(\varepsilon)\cdot\left[F_{\alpha}^{}(\tilde{q}_{1})-F_{\alpha}^{}(\tilde{q}_{2})\right]$ for any $\alpha$.  Therefore, when considering some finite $q_{2}$, discarding a subset of oblique incident angles from the noise background does not have an appreciable effect.  Still, in accordance with (13), the infinite-bandwidth limits are modified, strongly for $\alpha$ near or greater than 2, as 
\begin{equation}
P_{\alpha}(\eta,L,\varepsilon)\equiv\left(1-\varepsilon^{2-\alpha}\right)P_{\alpha}^{(0,\infty)}(\eta,L).
\end{equation}

Finally, we investigate narrow unimodal spectra where $G(q)$ has a maximum at some $q_{0}$.  First consider that if $G(q) \propto \delta(q-q_{0})$, then the resulting pressure is proportional to
\begin{equation}
P_{\delta}(\eta,\tilde{L}) = f(\tilde{L})+f(-\tilde{L}) = -\left.\frac{\partial F}{\partial \tilde{q}}\right|_{\tilde{q}=\tilde{L}}
\end{equation}
where $\tilde{L}\equiv q_{0}L$, with $f$ defined by Eq. (10) and $F$ defined by Eq. (3).  In the limit of zero dimensionless separation, 
\begin{equation*}
\lim_{\tilde{L}\to0}P_{\delta}(1,\tilde{L}) = -\frac{1}{3};
\end{equation*}
\begin{equation}
\lim_{\tilde{L}\to0}P_{\delta}(\eta,\tilde{L}) = \frac{2\eta}{3(1-\eta)}, \ \ \ \ \  \eta\neq1.
\end{equation}
$P_{\delta}(\tilde{L})$ is plotted in Fig. 5a for different values of $\eta$.  As $\tilde{L}$ is increased past an integer multiple of $\pi$, the force rapidly switches from attractive to repulsive, becoming discontinuous in the limit of perfect reflectors.  The cutoff $\tilde{q}_{c}(\eta)$ from Eqs. (5), (7) and Fig. 3a now gives the lowest value of $\tilde{L}$ at which $P_{\delta}(\tilde{L})$ is zero; when $\eta$ is near 1, $P_{\delta}$ quickly falls to a value slightly greater than $-1/3$ as $\tilde{L}$ is increased from $\tilde{q}_{c}$.  Below $\tilde{q}_{c}$, the force is repulsive, rapidly increasing in magnitude as $\tilde{L}$ is decreased further.  If one considers $q_{0}$ fixed, then $\tilde{q}_{c}$ is now a dimensionless critical separation, i.e. $L_{c} = \tilde{q}_{c}/q_{0}$.

We would like to see if this phenomena holds for general narrowly peaked spectra.  Taylor expanding an appropriate $G(q)$ about its maximum at $q_{0}$ gives a parabolic approximation [4],
\begin{equation*}
G(q) \approx G_{\text{par}}\max\left\{\frac{3}{4q_{0}D}\left[1-\left(\frac{q-q_{0}}{q_{0}D}\right)^{2}\right],0\right\}.
\end{equation*}
where $G_{\text{par}}$ is a constant energy density and the dimensionless parameter $D<1$ relates to the width of the peak.  This is equivalent to a weighted sum of three spectra, proportional to $1$, $q$, and $q^{2}$ respectively, over the narrow band $[q_{0}-q_{0}D,q_{0}+q_{0}D]$.  From Eq. (9), the resulting pressure is $G_{\text{par}}\text{Re}[P_{\text{par}}]$, where
\begin{equation}
P_{\text{par}}(D,\eta,\tilde{L})=\frac{3}{D^{3}}\left[\frac{D^{2}-1}{4q_{0}}P_{0}^{\{q_{0},D\}}(\eta,L)+\frac{1}{2q_{0}^{2}}P_{1}^{\{q_{0},D\}}(\eta,L)-\frac{1}{4q_{0}^{3}}P_{2}^{\{q_{0},D\}}(\eta,L)\right],
\end{equation}
and
\begin{equation*}
P_{\alpha}^{\{q_{0},D\}}(\eta,L)\equiv P_{\alpha}\left(\eta,L,q_{0}-q_{0}D,q_{0}+q_{0}D\right).
\end{equation*}
Note that $\tfrac{1}{G_{\text{par}}}\int_{0}^{\infty}dq\text{ }G(q) = 1$, and that the corresponding pressure depends on $q_{0}L$ and not separately on $q_{0}$ or $L$, cf. Eq. (11).  $P_{0}$ is just $P/I_{\omega}$ from Eqs. (2)-(3); a similar form of $P_{1}$, computationally more expedient than the result of (11)-(12), can be derived along the same lines of (2).  For $P_{2}$ this is not the case, but the form given by (11)-(12) is advantageous over numerical integration [6].

The result for $\eta=0.9$ is plotted in Fig. 5b for various $D$.  The phenomena near $\tilde{q}_{c}$ persists, and is in fact insensitive to the peak width!  When the bandwidth is fixed, both $P_{2}$ and $P_{1}$, like $P_{0}$, tend to large positive values as $L$ tends to zero, so long as the plates are not perfect reflectors; here, the repulsion at small $\tilde{L}$ is due to the positive $P_{1}$ term in Eq. (20), whereas the negative $P_{0}$ and $P_{2}$ terms counteract this tendency.  

The effect of larger widths comes into play at larger $\tilde{L}$, involving the fact that $n\pi(1-D)$ and $n\pi(1+D)$, $n$ an integer, become more spaced apart as $n$ increases; the alternations between attraction and repulsion become less rapid, and this effect is more pronounced for larger $D$.  It should be noted that the pressure resulting from a delta-spike, given by Eqs. (18) and (10), decays like $\tilde{L}^{-1}$; while Eq. (20) has the asympotic decay $\tilde{L}^{-2}$.  A nonzero peak width therefore mediates a transition between these two regimes.

\section*{Conclusions}

We have derived expressions for the pressure between two imperfectly reflecting plates immersed in random acoustic fields.  The asympotic properties of the associated special functions reveal singular behavior as the product of the plate reflectivities approaches unity, leaving a crucial imprint on the sign and magnitude of the force between closely-spaced plates in arbitrary spectral backgrounds.  The closer the plates are to perfect reflectors, the stronger the repulsion in the limit of no separation between the plates; while at the same time, increasingly small plate separations are required for attraction to cease.  It would be interesting to extend the analysis to include plate deformations due to the noise field [3] and prescribed spatiotemporal modulations of the plates [5,7].  

For completeness, we remark on the pressure between two plates with negative or complex $\eta$.  If plate A is a perfect reflector, $r_{A}=1$, and plate B is a pressure release surface, $r_{B}=-1$, then $\eta=-1$ and the wavevector component perpendicular to the plates can only take on values $(n\pi - \tfrac{\pi}{2})/L$, where $n$ is an integer.  The kinks, when $G(q)\propto q^{\alpha}$ over a finite band $[q_{1},q_{2}]$, and discontinuities, when $G(q) \propto \delta(q-q_{0})$, then occur at $Lq_{0,1,2} = n\pi - \tfrac{\pi}{2}$ rather than $n\pi$.  

If $G(q) \propto 1$, then the infinite-bandwidth limit is zero, in contrast to when $\eta=1$, but the $L\to0$ limit under finite bandwidth is the same as that of $\eta=1$.  Furthermore, if $G(q) \propto \delta(q-q_{0})$, then the $\tilde{L}\to0$ limit is $-1/3$, the same as when $\eta$ is strictly 1.  No singular behavior occurs as $\eta$ is increased from $-1$ (other than the smoothing of the kinks/discontinuities), and when $Lq_{0,2}$ is less than about $\tfrac{\pi}{2}$ the force is always attractive.  

The correspondence between $\delta$-spectra and parabolic spectra for positive $\eta$ also holds for negative as well as complex $\eta$.  If $\eta=e^{i\theta}$, there are kinks/discontinuities at $Lq_{0,1,2} = n\pi \pm \tfrac{\theta}{2}$, which smooth out as $\eta$ is varied along the radius of the unit circle.  Singular behavior can only occur when $\eta$ is varied to 1, cf. Eqs. (4), (14), and (19).  One can verify all this by evaluating Eqs. (2)-(3), (11)-(12), (18), (20) for any $\eta$ in the unit disk, taking only the real part of $P$ to get the desired acoustic Casimir force when $\text{Im}[\eta]\neq0$.    

\section*{References}

\noindent [1] H. B. G. Casimir, ‘‘\textit{On the attraction between two perfectly conducting plates},’’ Proc. K. Ned. Akad. Wet. 51, 793 

(1948).

\noindent [2] A. Larraza, C. D. Holmes, R. T. Susbilla, and B. Denardo, ‘‘\textit{The force between two parallel rigid plates due to the} 

\textit{radiation pressure of broadband noise: An acoustic Casimir effect},’’ J. Acoust. Soc. Am. 103, 2267 (1998).

\noindent [3] J. B\'arcenas, L. Reyes, and R. Esquivel-Sirvent, ``\textit{Acoustic Casimir pressure for arbitrary media},'' J. Acoust. Soc. 

Am. 116, 717 (2004).

\noindent [4] A. A. Lee, D. Vella, and J. S. Wettlaufer, ``\textit{Fluctuation spectra and force generation in nonequilibrium systems},'' 

Proc. Nat. Acad. Sci. 114, 9255 (2017).

\noindent [5] T. Emig, A. Hanke, R. Golestanian, and M. Kardar, ``\textit{Normal and lateral Casimir forces between deformed plates},'' 

Phys. Rev. A 67, 022114 (2003).

\noindent [6] $E_{m}(z)$ is given by the Wolfram Language function \texttt{ExpIntegralE}\textsf{[}\textsl{\textsf{m}}, \textsl{\textsf{z}}\textsf{].}  E. W. Weisstein, ``\textit{E\_n-Function},'' 

MathWorld-{}-A Wolfram Web Resource.  $<$https://mathworld.wolfram.com/En-Function.html$>$

\noindent [7]  A. Hanke, ``\textit{Non-Equilibrium Casimir Force between Vibrating Plates},'' PLoS ONE 8, e53228 (2013).

\end{document}